\documentclass[prl, onecolumn, showpacs, preprintnumbers, preprint]{revtex4}
\usepackage{graphicx,psfrag,citesort,dcolumn,bm,amsmath,amssymb}

\usepackage{textcomp} 
\newcommand{\Notiz}[1]{\ifdim\overfullrule>0pt\marginpar[{\raggedleft\textcolor[rgb]{1,0,0}{#1}}]{\raggedright\textcolor[rgb]{1,0,0}{#1}}{}\fi} 

\begin{document}

\title{Transient Binding and Dissipation in Semi-flexible Polymer Networks}

\author{O. Lieleg$^1$, M.M.A.E. Claessens$^{1,2}$, Y. Luan$^{1,3}$ and A. R.  Bausch$^1$} 
\affiliation{$^1$Lehrstuhl f\"ur Biophysik E22, Technische Universit\"at M\"unchen, Germany, $^2$MESA , Faculty of Science and Technology, University of Twente, The Netherlands, $^3$School of Pharmacy, Shandong University, P.R. China}

\date{\today}

\begin{abstract}
While polymer solutions lack the mechanical stability only transiently cross-linked networks can fulfill the competing requirements of structural stability and maximal energy dissipation. Here, we show that transient cross-links entail local stress relaxation and energy dissipation in an intermediate elasticity dominated frequency regime. We quantify the mechanical response of a semi-flexible polymer network by experimentally tuning the off-rate of the transient cross-linker molecule and theoretically reproduce the measured frequency response by a model that is predicated on microscopic unbinding events. \end{abstract}


\maketitle


In polymer networks distinct molecular mechanisms can lead to a relaxation of an external (or internal) stress on different time scales; depending on whether energy is stored or dissipated, elastic or viscous behavior can be evoked. Many synthetic polymer networks are covalently cross-linked resulting in a predominantly elastic response over a broad frequency range. For covalently cross-linked polymer networks the molecular origin of the viscoelastic response is well understood: viscous drag (Rouse friction) resulting from bending undulations of single polymers and internal conformational changes (Kuhn friction) play an important role~\cite{DeGennes1979}, the covalent cross-links suppress single polymer diffusion. Therefore, the frequency dependent viscoelastic behavior of a permanently cross-linked polymer network is expected to resemble that of a damped harmonic oscillator and should reach a constant level of elasticity at low frequencies while viscous effects become negligible. However, the experimentally observed frequency response for various soft materials like worm-like micelles~\cite{Cates1990, Won1999, Buchanan2005}, equilibrium polymers~\cite{Sjibesma1997, vanderGucht2003}, biopolymer networks and living cells~\cite{Dheng2006, Massiera2007} is in marked contrast to this theoretical expectation. In spite of the presence of physical cross-links based on electrostatic interactions or van-der-Waals forces, a pronounced minimum and maximum in the viscous dissipation is observed for these materials in a frequency range of $0.01~-~10$~Hz. This feature is always accompanied by a decrease in elasticity for low frequencies; nevertheless, the viscoelastic response in this frequency regime is dominated by the network elasticity. For the semi-flexible polymers used by cells it is of utmost importance not only to withstand mechanical strains but also to allow for continuous remodeling of the cellular microstructure which can only be achieved by transient cross-links. It is yet to be resolved how the unique properties of this class of transient cross-links affect the frequency response of cross-linked polymer networks on an intermediate frequency regime dominated by the network elasticity. Transient cross-links can be characterized by an off-rate, $k_{\mathrm{off}}$, which typically corresponds to frequencies that are on the order of several mHz up to a few Hz. Hitherto, research focused on the modification of the viscoelastic response in both asymptotic frequency regimes: The ''stickiness'' of cross-linking molecules or polymer sidegroups gives rise to additional friction processes. In soft materials the sticky Rouse relaxation at high frequencies ($f~>>~k_{\mathrm{off}}$) as well as the slowed-down sticky reptation process ($f~<<~k_{\mathrm{off}}$) have already successfully been described~\cite{Leibler1991, Tanaka1992, Rubinstein2001}. However, the molecular mechanisms responsible for the behavior at intermediate frequencies ($f~\approx~k_{\mathrm{off}}$) are poorly understood -- yet urgently needed for a molecular understanding of the mechanical properties of soft materials, including biomaterials and living cells. 

Here, we show that unbinding of transient cross-links results in a stress release mechanism in cross-linked semi-flexible polymer networks. This stress release mechanism decreases the static network elasticity and at the same time increases the viscous dissipation in the network. The timescale of this stress release is set by the lifetime distribution of the cross-linking molecules and can therefore be tuned independently from Rouse-like friction. To quantitatively explore these transient binding effects, we make use of a well-defined model system, actin filaments that are cross-linked by rigor-HMM, which exhibits transient cross-links characterized by a single unbinding rate. 


G-actin is obtained from rabbit skeletal muscle following~\cite{Spudich1971}, stored and polymerized into filaments as described before~\cite{Lieleg2007, Lieleg2007a}. HMM is prepared from Myosin II by chymotrypsin digestion and tested using motility assays as in~\cite{Uhde2004}. In the experiments the molar ratio $R$ between HMM and actin, $R = c_{\mathrm{HMM}}/c_\mathrm{a}$, is varied. The formation of cross-linked rigor networks is recorded as described in~\cite{Tharmann2007}, the viscoelastic response is recorded in the linear regime as described in~\cite{Lieleg2007, Lieleg2007a}. $\gamma$-ATP (AMP$\cdot$P-N-P) or glycerol is added before the actin polymerization is initiated; ATP depletion is recorded, accordingly. Glutaraldehyde is added together with HMM to pre-polymerized actin filaments. The resulting solution is gently mixed with a cut pipette tip to avoid breaking of filaments and then loaded into the rheometer (Physica MCR 301, Anton Paar, Graz, Austria) for structural equilibration and ATP depletion.


Proteins such as fascin~\cite{Lieleg2007}, $\alpha$-actinin~\cite{Wachsstock1994}, filamin~\cite{Gardel2006} or rigor-HMM~\cite{Tharmann2007} create non-covalent cross-links in actin networks. In these networks, the viscous dissipation ($G''(f)$) exhibits a pronounced minimum at a frequency $f_{\mathrm{min}}$, which exact position depends on the cross-link density (Fig.~1). As an example, the elastic response of cross-linked actin/rigor-HMM networks differs from the expectation for covalently cross-linked networks as the elastic modulus ($G'(f)$) is only constant for sufficiently large frequencies $f~>~f_{\mathrm{min}}$ but decreases significantly at lower frequencies. In this low frequency regime, the viscous dissipation reaches a maximum around $f_{\mathrm{max}}~\sim~0.03$~Hz. This time scale is independent of the cross-link density; however, the maximal amount of low-frequency energy dissipation increases linearly with the cross-link density. 

\begin{figure}[h!]
\includegraphics[width = 0.6\columnwidth]{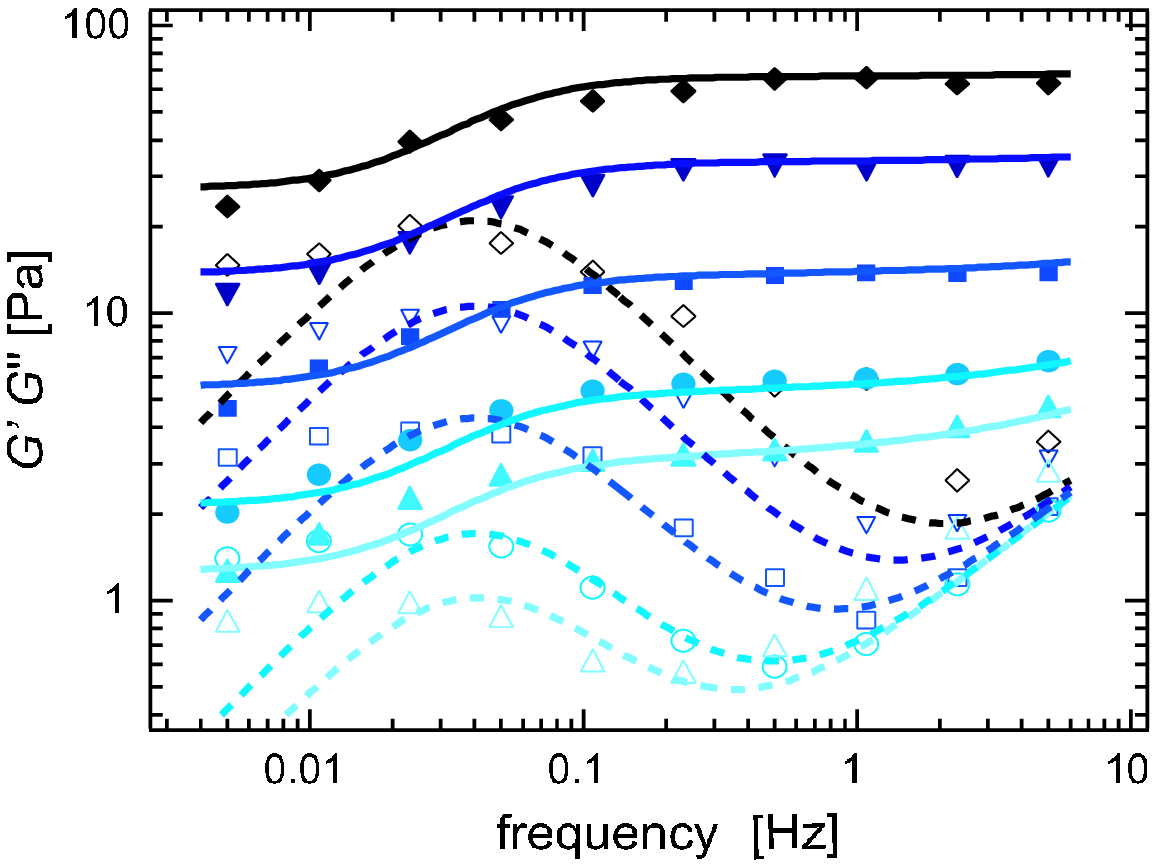}
     \caption{Elastic (full symbols) and viscous (open symbols) response for actin/rigor-HMM networks as a function of frequency ($c_a = 19 ~\mu\mathrm{M}$, $R$~=~0.0076 (upright triangles) up to $R$~=~0.143 (diamonds)). The full and dashed lines represent a global best fit of the model described in the main text. All parameters are constant with the exception of the cross-link density $N$. Here, $N~\sim~R^{1.1}$ is used in excellent agreement with the experimental finding $G_0~\sim~R^{1.2}$ (suppl. inform.).}
          \label{FIG1}
\end{figure}

To shed light on the molecular origin of this viscoelastic response the possible molecular mechanisms involved should be tuned independently. A mechanism of energy dissipation that is always present in polymer networks is Rouse friction due to viscous drag of individual filaments. The viscosity of the solvent can be increased by the addition of glycerol. As a consequence, the time scale of the Rouse relaxation should be shifted according to the increase in solvent viscosity. Indeed, for the cross-linked network the minimum in the viscous dissipation is relocated as the viscosity of the solvent is increased by glycerol~(Fig.~2A). However, the addition of glycerol does not affect the elastic network response over almost the whole frequency range. 

It is important to note, that the viscous dissipation depends on the solvent viscosity only at frequencies $f~>~f_{\mathrm{min}}$ while e.g. the maximal dissipation at $f_{\mathrm{max}}~\approx~0.03$~Hz remains unchanged. This agrees with previous results~\cite{Tharmann2007, Lieleg2007a} which also indicated that Rouse friction is not sufficient to rationalize the complex dissipation behavior of a cross-linked polymer network. In fact, another molecular mechanism has to be considered. Recall that the maximum in the viscous dissipation is located around $f_{\mathrm{max}}~\approx~0.03$~Hz independent of the cross-link density (Fig.~1B). The cross-links are created by the protein HMM which is known to form a non-covalent bond to the biopolymer actin with a typical off-rate $k_{\mathrm{off}}~\approx~0.09~\mathrm{s}^{-1}$~\cite{Marston1982}. As this off-rate is on the order of $f_{\mathrm{max}}$, the binding kinetics of the cross-linking protein HMM might give rise to additional mechanisms in the network accounting for the observed frequency dependence of the viscoelastic moduli for frequencies $f~<~f_{\mathrm{min}}$.

It was shown before that the bond between actin and actin binding proteins (ABPs) can be forced to unbind in the presence of mechanical forces~\cite{Guo2006, Miyata1996}. However, the transient nature of an actin/ABP bond should also allow for spontaneous unbinding events in thermal equilibrium. Thus, thermal unbinding of distinct cross-links could be the molecular reason for the observed behavior of both viscoelastic moduli in the linear response regime.
To verify this hypothesis a fixation of the ABP/actin bond would be of avail. Glutaraldehyde is able to create a covalent linkage between neigboring molecules by the formation of a Schiff base. It is commonly used for fixation purposes of cells and other biomaterials~\cite{Fernandez2007} and can be employed to create a chemical bond between HMM and actin. Indeed, as depicted in Fig.~2B, the minimum in the viscous dissipation can be suppressed by the addition of 0.1 $\%$ glutaraldehyde. At the same time, the decrease in the network elasticity at low frequencies as observed in the absence of glutaraldehyde is almost completely suppressed. This suggests that the minimum in the viscous dissipation marks the timescale at which the transient character of the cross-links starts to dictate the viscoelastic response of the network.

\begin{figure}[h!]
\includegraphics[width = 0.43\columnwidth]{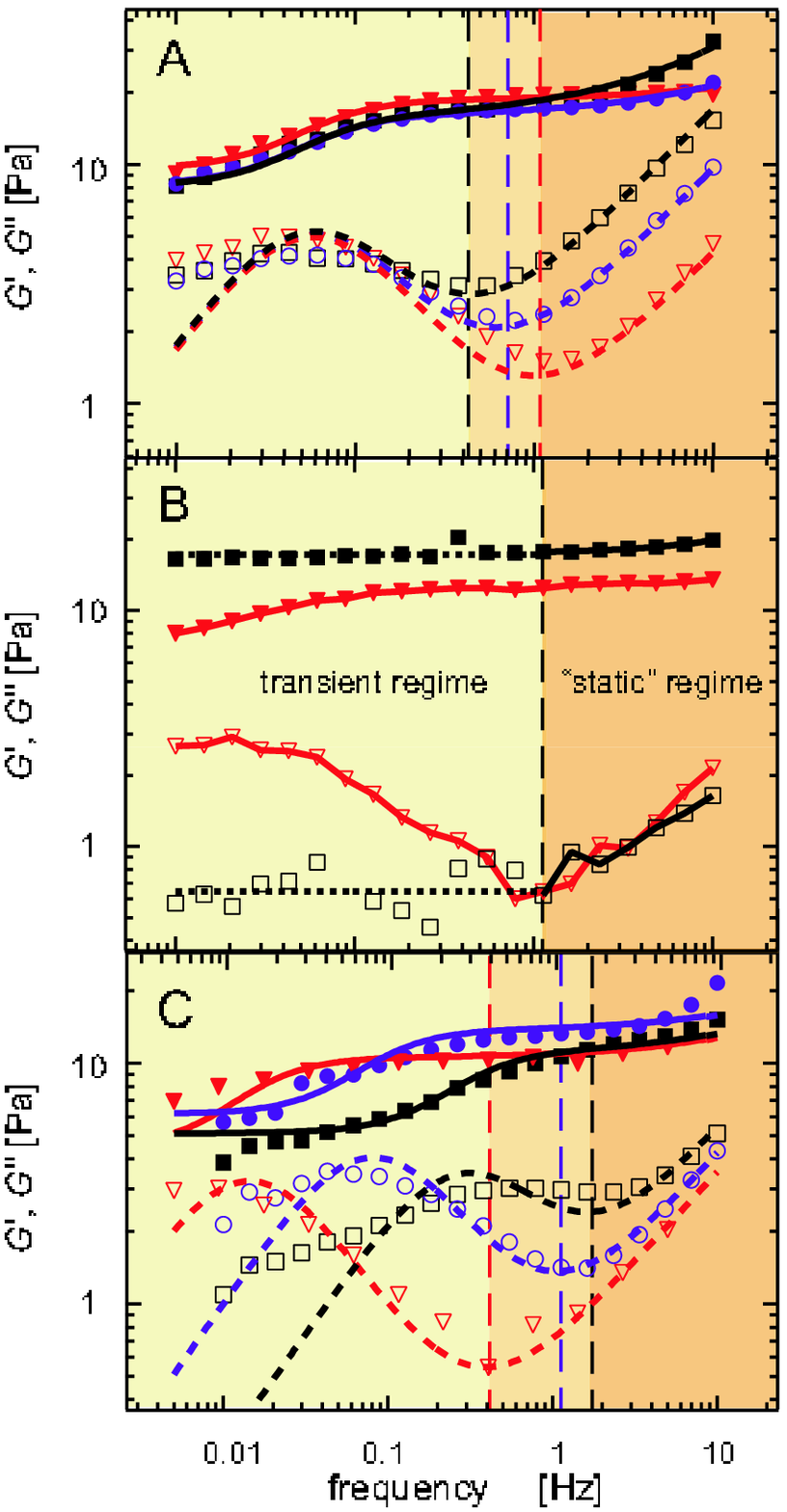}
	\caption{Elastic (full symbols) and viscous (open symbols) response for actin/HMM networks ($c_a = 9.5~\mu\mathrm{M}$, $R~=~0.1$). The full and dashed lines in (A) and (C) represent a global best fit of the model as discussed in the main text. Parameters are adjusted as they are controlled experimentally (see suppl. inf.) (A) Distinct amounts of glycerol ($0~\%$ (triangles), $25~\%$ (circles) and $50~\%$ (squares) are added to increase the solvent viscosity. (B) $0.1~\%$ glutaraldehyde is added to fix the actin/HMM bond. The resulting viscoelastic response (squares) is compared to a non-fixed network (circles). The dotted lines are constant fits to guide the eyes. (C) Distinct amounts of $\gamma$-ATP (0~mM (triangles), $50~\mu\mathrm{M}$ (circles) and 2~mM (squares)) are added to tune the off-rate of the actin/HMM bond. The parameter $R$ is adjusted according to the decrease in HMM affinity to obtain networks with comparable static elasticity $G_0$. (A), (B), (C): The vertical lines represent the transition from the transient regime to the ''static'' regime as described in the main text. 
	}
          \label{FIG2}
\end{figure}

The covalent fixation of an actin/ABP bond using glutaraldehyde has an extreme effect on the viscous dissipation. A more subtle method to affect the viscous response of a cross-linked polymer network might be given by only slightly changing the binding kinetics of the cross-linking molecule. It was shown, that $2~\mathrm{mM}$ $\gamma$-ATP, a non-hydrolizable ATP-analogon, can increase the off-rate of one actin/HMM bond from $0.09~\mathrm{s}^{-1}$ up to $1.8~\mathrm{s}^{-1}$~\cite{Marston1982}. To quantitatively analyse how altered binding kinetics of the cross-linking molecule influence the viscoelastic response, networks with different amounts of $\gamma$-ATP but comparable network elasticity and thus with comparable equilibrium cross-link densities are investigated. First, the addition of $\gamma$-ATP to a cross-linked actin/HMM network changes the position of both the minimum and the maximum of the viscous dissipation (Fig.~2C). However, for frequencies $f~>~f_{\mathrm{min}}$ the viscoelastic response seems almost unaffected. Second, the frequency at which the network elasticity starts to drop is also shifted with the increase in $k_{\mathrm{off}}$. This strongly suggests that these two features have the same molecular origin.

The results presented so far imply that thermal unbinding of distinct cross-links triggers a relaxation mechanism which influences the elastic and the viscous properties of the network simultaneously. In thermal equilibrium, an ensemble of $N$ cross-links exhibits statistical unbinding events whose probability is determined by the cross-linker off-rate, $k_{\mathrm{off}}$. This unbinding can be described in analogy to a unimolecular reaction: $\mathrm{AB}\stackrel{k_{\mathrm{off}}}{\longrightarrow} \mathrm{A + B}$. Herein, the rebinding process is assumed to be fast enough to provide a constant
equilibrium number of cross-links: the unbinding process is only limited by the lifetime of the actin/ABP bond and not by the reformation of new bonds. This results in an exponential distribution of unbinding events in the time domain,
$t~\geq~0$: $N(t) \sim N \cdot e^{-k_{\mathrm{off}}\cdot t}$ which can be translated into the frequency domain using a Fourier transformation yielding the complex function $\widehat{N}(f)$. Then, the real part $\Re(\widehat{N}(f))$ represents the loss of elasticity due to cross-link unbinding - the static network elasticity $G_0$ is reduced:
\begin{equation}
        G'(f) = G_0 - a\cdot\frac{Nk_{\mathrm{off}}}{\frac{k_{\mathrm{off}}^2}{4\pi^2}+f^2} + b\cdot\left(\frac{f}{f_0}\right)^{3/4}
            \label{modelG'}
    \end{equation}
where the last term represents the Rouse relaxation of single filaments in semi-flexible polymer networks~\cite{Morse1998, Gisler1999}. The time scale of the Rouse mode is set by the factor $f_0$ which is a function of the solvent viscosity $\eta$~\cite{Rubinstein2003}. $\Im(\widehat{N}(f))$ contributes to the viscous part of the frequency spectrum
where it competes with the dissipative part of the Rouse relaxation:
    \begin{equation}
        G''(f) = c\cdot\frac{Nf}{\frac{k_{\mathrm{off}}^2}{4\pi^2}+f^2} + d\cdot \left(\frac{f}{f_0}\right)^{3/4}
            \label{modelG''}
    \end{equation}
The key parameters are the cross-linker off-rate which is known from independent experiments~\cite{Marston1982} and the number of cross-linking molecules; the prefactors $a$ and $c$ include the amount of energy that is released and dissipated 
by unbinding of a single cross-link, $b$ and $d$ depend on the density of filaments in the network which is kept constant during a set of measurements. The best fit for the data set shown in Fig.~1 is obtained for $k_{\mathrm{off}}~\approx~0.3~\mathrm{s}^{-1}$, which is in excellent agreement with values determined by biochemical means~\cite{Marston1982}.

Not only the maximal amount of dissipated energy, $G''(f_{\mathrm{max}})$, is quantitatively reproduced but also the minimum in the viscous dissipation and their dependencies on $k_{\mathrm{off}}$ and $N$. The minimum in the viscous dissipation can thus be identified as a direct result of the competition between local stress release triggered by cross-link unbinding and Rouse friction of single filaments. Importantly, the decay in the elastic response is also correctly reflected by the global fit. It can be seen, that the mechanism of local stress release becomes increasingly important when approaching $f~\approx~\frac{k_{\mathrm{off}}}{2\pi}$ where the contribution of unbinding events, $\widehat{N}(f)$, is most pronounced and the viscous dissipation becomes maximal. If the off-rate of the cross-linking molecule is altered (Fig.~2C), the maximum in the viscous dissipation is shifted, accordingly. Indeed, the fits were obtained using literature values for $k_{\mathrm{off}}$ at distinct amounts of $\gamma$-ATP as determined by independent measurements~\cite{Marston1982}. 
In all experiments, at very high frequencies $f~>>~\frac{k_{\mathrm{off}}}{2\pi}$ the viscous response is unaffected by the stress release mediated by unbinding events. Vice versa, a variation of the solvent viscosity results only in a shift of the Rouse relaxation time without affecting the off-rate of the cross-linking molecules. Consistently, only the known solvent viscosity needs to be considered to reproduce the viscoelastic network response in the presence of glycerol (Fig.~2A).

It is important to note, that the observed transient unbinding effects described here do not permanently change the material properties -- underlining the thermal nature of the process.  
The transient binding discussed here might be a mechanism that provides maximal energy dissipation without permanently altering the microstructure of the network. The tunability of the viscous dissipation in an elastic dominated frequency regime allows for local reorganization processes and creates an adaptive material which is able to absorbe mechanical shocks on the microscopic scale without causing structural failure. In fact, this extraordinary material property might be an important advantage for living cells employing transiently cross-linked biopolymer networks instead of covalently cross-linked structures which would be much too brittle.

We have shown that the viscous dissipation in biopolymer networks does not follow theoretical expectations for a covalently cross-linked polymer network. The viscous dissipation occurring in such materials on an intermediate time scale can be rationalized by a molecular stress release mechanism based on transient cross-linker unbinding. This mechanism superimposes dissipation mechanisms that are already known for polymer networks and solutions. The frequency at which this mechanism of stress release dominates over single polymer friction is set by the off-rate of the cross-linking molecule, the cross-link density and the viscosity of the solvent. This defines a mechanical transition regime: If the system is deformed with a frequency much faster than $f_{\mathrm{min}}$, cross-linker unbinding is too slow to significantly modify the viscoelastic response of the network, the cross-links appear to be ''static'' and unbinding kinetics are more or less irrelevant (Fig.~2B). However, if the deformation is imposed slow enough, the transient nature of the cross-links will dictate both the elastic and the viscous response of the network. Additionally, the semi-flexible nature of the polymer itself in combination with unspecific intermolecular interactions gives rise to a broad range of slow dynamics as observed for entangled actin solutions~\cite{Semmrich2007}. One might also need to account for this ''glassy'' behavior~\cite{KroyGWLC} in order to rationalize the viscoelastic response of a cross-linked semi-flexible polymer network over the full frequency range. 

The amount of energy which is dissipated in molecular unbinding and the ensuing stress release is determined by the density and the lifetime distribution of the cross-links. The tunability of these parameters provides adaptability in biomaterials; they can modulate their elastic and viscous properties without any permanent structural change. The transient binding and energy dissipation discussed here is indispensible for a detailed understanding and further development of adaptable biomaterials. Moreover, transient binding effects might also be employed by cells for mechanosensing tasks that take place on time scales comparable to the frequency range investigated in this study -- especially since the non-static nature of actin/ABP bonds gives rise to a high sensitivity towards external or internal forces.

\small{We thank M. Rusp for the actin preparation. Financial support via the ''Munich-Centre for Advanced Photonics (MAP)'' is gratefully acknowledged. O. Lieleg acknowledges support from CompInt.}

\newpage

\section*{Supplementary information for Lieleg et al.}

\subsection*{Dependence of the static network elasticity on the cross-link density}

The static network elasticity $G_0$ of an isotropically cross-linked semi-flexible polymer network can be described by an affine stretching model~[S1] as
\begin{equation}
		G_0 \sim \frac{\kappa_0^2}{k_\mathrm{B} T \xi^2 l_\mathrm{c}^3}
	\label{G0streck}
\end{equation}

\noindent where $\kappa_0$ denotes the bending modulus of a single actin filament, $\xi$ is the meshsize of the network and $l_c$ describes the average distance between two cross-links. Hence, the term $\frac{1}{l_\mathrm{c}^3}$ gives the density of cross-links which is proportional to the number of cross-link points, $N$.
Thus, equation (\ref{G0streck}) can be rewritten to 
\begin{equation}
		G_0 \sim \frac{\kappa_0^2\cdot N}{k_\mathrm{B} T \xi^2}
	\label{G0modified}
\end{equation}

\noindent which reduces to the simple scaling argument $G_0 \sim N$ valid for constant bending modulus and mesh size. The former condition is fulfilled as for the actin/HMM system bundling of filaments does not occur. As a consequence, the mesh size of actin/HMM networks is also independent of the cross-link density as long as the filament density is kept constant (which is the case for the data presented in Fig.~1).

\subsection*{Fitting parameters}

\paragraph*{Stress relaxation coefficients:}
For all data sets constant values were used for the prefactors of the unbinding contribution, i.e. $a~=~3\cdot10^{-16}~\mathrm{Pa}\cdot \mathrm{s}^{-1}$ and $c~=~2\cdot 10^{-15}~\mathrm{Pa}\cdot \mathrm{s}^{-1}$.

\paragraph*{Rouse relaxation coefficient:}
To model the Rouse friction in the absence of glycerol $b~=~0.4~\mathrm{Pa}\cdot \mathrm{s}^{3/4}$ and $d~=~0.7~\mathrm{Pa}\cdot\mathrm{s}^{3/4}$ is used in Fig.~1 and Fig.~2C. By the addition of glycerol (Fig.~2A) the viscosity of the solvent $\eta$ is increased up to 6-fold for 50~\% glycerol~[S2]. The time scale of the Rouse relaxation can be shifted as a scaling factor $f_0$ can be introduced in the last term of equation (1) and (2) of the manuscript: $G''(f)_{Rouse}~\sim \left(\frac{f}{f_0}\right)^{3/4}$. From the resulting apparent increase of $b$ and $d$ with the solvent viscosity a scaling $f_0~\sim~1/\eta$ can be obtained as expected~[25].

\paragraph*{Off-rate:}
For the data shown in Fig.~1 and Fig.~2A the best fits are obtained for an off-rate of $k_{\mathrm{off}}~=~(0.3~\pm~0.05)~\mathrm{s}^{-1}$ which is in good agreement with the literature value $k_{\mathrm{off}}~=~0.09~\mathrm{s}^{-1}$~[19] which is the off-rate for a single HMM-head in its ADP-rigor state.
In Fig.~2C literature values as determined by \cite{Marston1982} were used, i.e. $k_{\mathrm{off}}(0~\mathrm{mM}~\gamma\mathrm{ATP})~=~0.09~\mathrm{s}^{-1}$ and $k_{\mathrm{off}}(2~\mathrm{mM}~\gamma\mathrm{ATP})~=~1.8~\mathrm{s}^{-1}$.

Note, that $\gamma$-ATP does not only incrase the off-rate of the actin/HMM bond but also decreases the binding constant of HMM towards actin.

\vspace{1 cm}
\noindent\protect[S1] F. C. MacKintosh \textit{et al.}, Phys. Rev. Lett. \textbf{75}, 4425 (1995)\\
\protect[S2] N. E. Dorsey, \textit{Properties of Ordinary Water-Substance} (Hafner Publishing Co., New York, 1940)

\end{document}